\documentclass[prl,twocolumn,showpacs,floatfix]{revtex4-1}
\usepackage{graphics,epsfig,amsfonts,amssymb,amsmath,soul}
\begin{document}
\title{Inter-event correlations from avalanches hiding below the detection threshold}
\author{Sanja Jani\'{c}evi\'{c}$^{1}$}
\author{Lasse Laurson$^{1,2}$}
\email{lasse.laurson@aalto.fi}
\author{Knut J\o rgen  M\aa l\o y$^3$}
\author{St\'ephane Santucci$^{3,4}$}
\author{Mikko J. Alava$^1$}
\affiliation{$^1$COMP Centre of Excellence, Department of Applied 
Physics, Aalto University, P.O. Box 11100, 00076 Aalto,
Espoo, Finland}
\affiliation{$^2$Helsinki Institute of Physics, Department of Applied
Physics, Aalto University, P.O. Box 11100, 00076 Aalto,
Espoo, Finland}
\affiliation{$^3$Department of Physics, University of Oslo, PB 1048
Blindern, NO-0316, Norway}
\affiliation{$^4$Laboratoire de physique, CNRS UMR 5672, Ecole
Normale Sup\'erieure de Lyon, 46 All\'ee d'Italie, 69364 Lyon
Cedex 07, France}

\begin{abstract}
Numerous systems ranging from deformation of materials to earthquakes
exhibit bursty dynamics, which consist of a sequence of events with a broad event size
distribution. Very often these events are observed to be temporally correlated or 
clustered, evidenced by power-law distributed waiting times separating two 
consecutive activity bursts. We show how such 
inter-event correlations arise simply because of a finite detection threshold, 
created by the limited sensitivity of the measurement apparatus, or used 
to subtract background activity or noise from the activity signal. 
Data from crack propagation experiments and numerical simulations of a non-equilibrium 
crack line model demonstrate how thresholding 
leads to correlated bursts of activity by separating the avalanche events into 
sub-avalanches. The resulting temporal sub-avalanche correlations are well-described 
by our general scaling description of thresholding-induced correlations in crackling 
noise.
\end{abstract}

\maketitle

A large class of physical, biological and other systems respond 
to slowly changing external conditions 
by exhibiting scale-free avalanche dynamics, or ``crackling noise'' \cite{SET-01},
measurable as a bursty activity signal $V(t)$. 
Depending on the system, $V(t)$ may originate from a number of processes:
the velocity of a propagating crack \cite{SCH-97,MAL-06,LAU-13,LAU-10} or the plastic
deformation rate \cite{ALA-14,ZAI-06,MIG-01,DIM-06,CSI-07} in a
stressed solid, the fluid invasion rate into porous media \cite{ROS-07,SAN-11},
the rate of change of magnetization in a dirty ferromagnet
in a slowly changing external magnetic field \cite{DUR-06,DUR-00},
or time-dependent activity in neuronal networks \cite{BEG-03,BEL-15}.
In many cases, the critical-like scaling implied by the power-law burst size distributions
has found an interpretation in terms of a non-equilibrium phase
transition \cite{LUB-04}, separating quiescent and active phases of the
system \cite{FIS-98}, and making it possible to apply concepts and tools such as
universality and renormalization group theory \cite{DOU-09}.

Another key feature of typical crackling noise signals is
that the bursts often exhibit temporal correlations,
visible as power-law distributed waiting times (quiet times, or periods
of low activity) separating two consecutive events \cite{BEN-06,SAL-02,STO-14,
TAN-13,MAK-15,BAR-13,PLE-09};
in contrast to these observations, uncorrelated triggering of avalanches would
be described by a Poisson process, with exponentially distributed
waiting times. The perhaps best-known example of such temporal
correlations is the spatio-temporal clustering of earthquakes \cite{BEN-06},
often described by phenomenological laws like the Omori law \cite{OMO-94,DAR-16}.
Similar time-clustering of events or power-law distributed waiting times
are also observed in acoustic \cite{SAL-02,STO-14} and light \cite{TAN-13} 
emission from fracture, compression of wood samples \cite{MAK-15} and 
porous materials \cite{BAR-13}, as well as for neuronal avalanches 
\cite{PLE-09}.

From a theoretical perspective, the typical quasistatically driven
model systems (of propagating cracks, invasion fronts, domain walls, etc.) where the 
bursty activity stems from an underlying dynamical phase transition 
fail to reproduce the empirically observed
strong temporal inter-event correlations, thus raising the question of their origin.
If one incorporates additional slow processes \cite{PAP-12} (e.g., viscoelasticity
\cite{JAG-14}) in these models, temporal avalanche clustering may be recovered.
However, such attempts merely call for more general explanations of the empirical
observations of inter-event temporal correlations in a variety of crackling noise
systems.

By using experimental data from planar crack propagation experiments and numerical 
simulations of a crack line model, we show how temporal avalanche correlations in
crackling noise simply result from the thresholding process used to define the 
bursts or avalanches \cite{LAU-09,FON-15}. This thresholding is often necessary: 
it is applied either indirectly (due to a finite detection threshold or sensitivity
of the experimental apparatus) or actively (when finite activity background or 
noise level needs to be subtracted from $V(t)$ to look for avalanches).
The full avalanche events -- which are correlated sequences of
activity by definition -- are partly ``hiding'' below the finite detection 
threshold, and thus broken into sub-avalanches in the thresholding process.
This leads to correlations between the observed events, even if the underlying
``true'' avalanche triggers can be well-described by a Poisson process. We 
present a general scaling description of the thresholding-induced
(sub-)avalanche correlations, 
and find that our experimental 
and numerical results are in excellent agreement with the resulting predictions. 

\begin{figure}[t!]
\includegraphics[trim=0.25cm 1.25cm 0cm 0.75cm, clip=true, width=8.0cm]{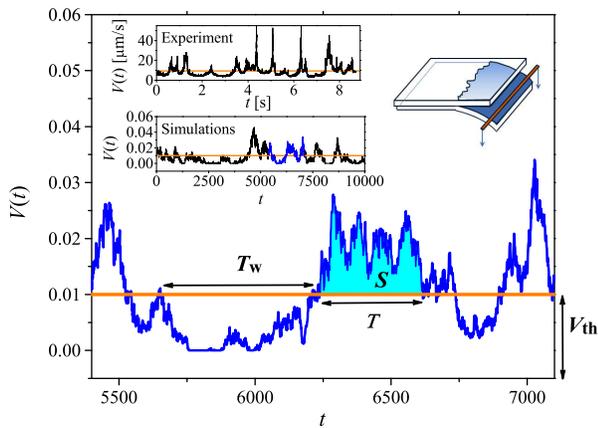}
\caption{Insets on the left show examples of the experimental (top) and numerical
(bottom) crack front velocity time series $V(t)$. The blue part
of the latter is shown magnified in the main figure, with definitions
of the avalanche size $S$, duration $T$ and waiting time $T_\text{W}$,
resulting from applying a finite threshold level $V_\text{th}$ (orange
line). The geometry of the experiment is portrayed in the top right
inset.\label{fig:1}}
\end{figure}

When defining bursts or avalanches from a bursty signal $V(t)$ by
thresholding, a finite threshold level $V_\text{th}$ is imposed, and
excursions of $V(t)$ above $V_\text{th}$ are identified as  events of
interest, see Fig. \ref{fig:1}. Their sizes
$S = \int_0^{T}\text{d}t [V(t)-V_\text{th}]$ and durations $T$ are
power-law distributed with a cutoff, that is, $P(S)=S^{-\tau_S}f(S/S_0)$ and
$P(T)=T^{-\tau_T}g(T/T_0)$, with $f(x)$ and $g(x)$ scaling functions,
and $S_0$ and $T_0$ the cutoff avalanche size and duration, respectively.
The average avalanche size scales with the duration as
$\langle S(T) \rangle \propto T^{\gamma}$, with the critical exponents
expected to satisfy the scaling relation $\gamma = (\tau_T-1)/(\tau_S-1)$.
The average burst amplitude would then scale as $T^{\gamma-1}$ \cite{LAU-13}.

\begin{figure}[t!]
\includegraphics[width=7.5cm]{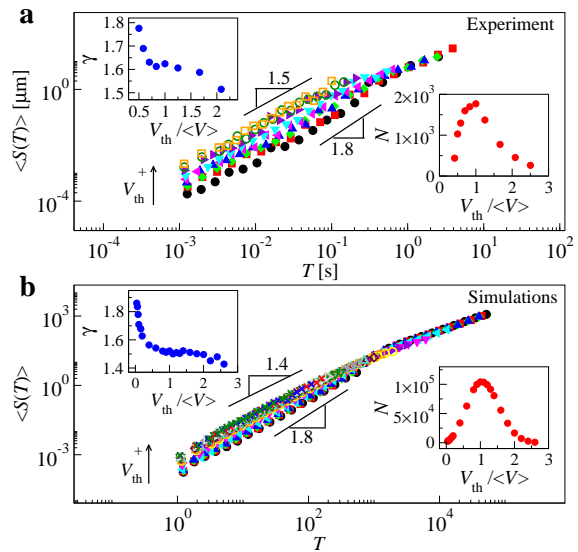}
\caption{Scaling of $\langle S(T) \rangle$ with $T$ for a wide range of threshold levels 
$V_\text{th}$, with arrows indicating the direction of rising $V_\text{th}$. The experimental
data for $\langle V \rangle$ = 10.2 $\mu$m/s and for $V_\text{th}$ varied in the
range 5.1 - 25.5 $\mu$m/s are shown in {\bf a}, with the
corresponding numerical results for $\langle V \rangle = 0.025$ and $V_\text{th}$
in the range 0.001-0.065 in {\bf b}. 
The top left insets show the evolution of the effective value of 
$\gamma$ with $V_\text{th}$, resulting from a fit to the scaling range
of the $\langle S(T) \rangle$ data.
The bottom right insets display the
$V_\text{th}$-dependence of the number of (sub)avalanches, exhibiting a maximum
at $V_\text{th} \approx \langle V \rangle$.\label{fig:2}}
\end{figure}

\begin{figure*}[t!]
\begin{center}
\includegraphics[trim=0cm 1.8cm 0cm 0cm, clip=true,width=15cm]{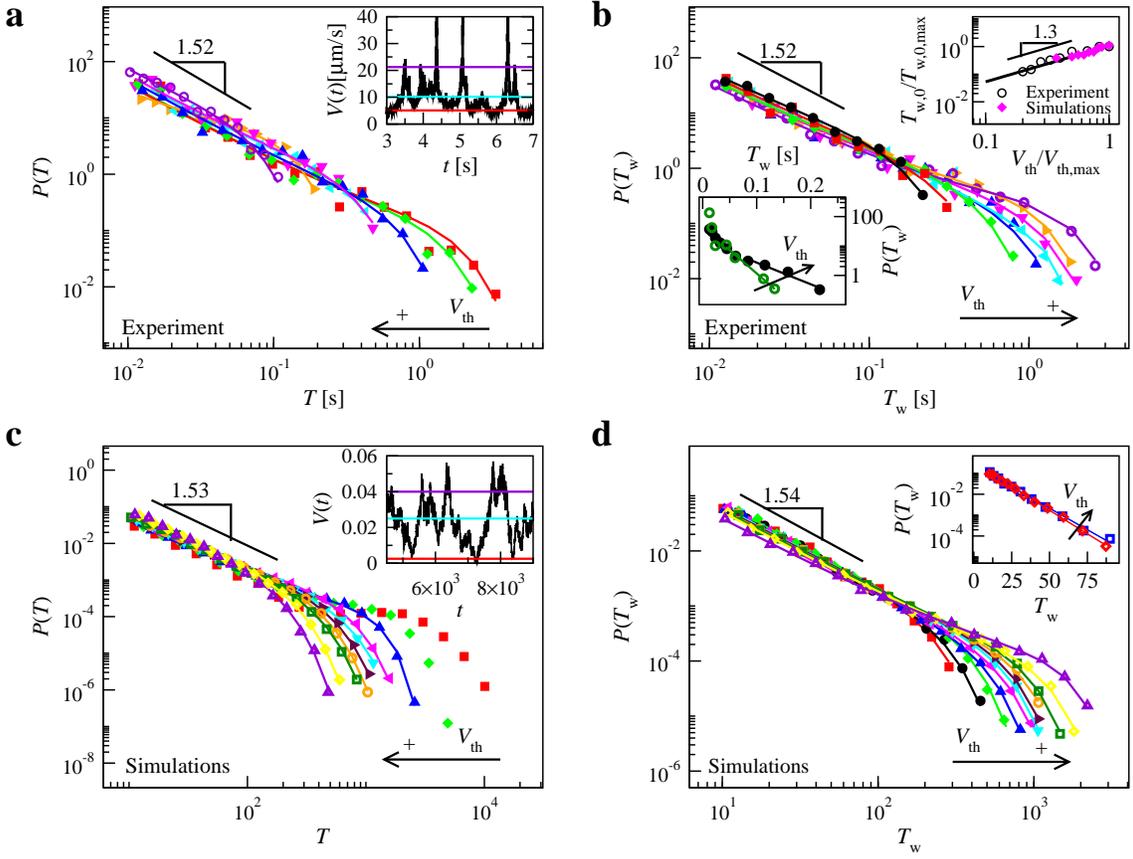}
\caption{The main panels of {\bf a} and {\bf b} show experimental 
$P(T)$ and $P(T_\text{W})$ distributions, respectively, for
$\langle V \rangle$ = 10.2 $\mu$m/s and a wide range of 
$V_\text{th}$'s, with arrows indicating the direction of rising $V_\text{th}$; 
the corresponding data from simulations           
are shown in {\bf c} and {\bf d}, with $\langle V \rangle = 0.025$.
The horizontal lines in the insets of {\bf a} and {\bf c} illustrate the
maximum and minimum $V_\text{th}$ used, as well as 
$\langle V \rangle$, showing a typical part of the $V(t)$ signal for
reference. $P(T_\text{W})$ evolves from an exponential
(bottom left inset of {\bf b} and the inset of {\bf d} show examples also for
even smaller $V_\text{th}$s than in the main panel(s) on a semilog scale)     
to a power law with a cutoff as $V_\text{th}$ is increased; 
$\tau_{T_\text{W}}$ equals $\tau_T = 1.53 \pm 0.05$ within
errorbars. The cutoff $T_\text{W,0}$ of the waiting time distributions
grows with $V_\text{th}$ as $T_\text{W,0} \propto V_\text{th}^{\delta}$,
with $\delta \approx 1.3$ (top right inset of {\bf b}, showing data from both
simulations and experiment)\label{fig:3}. Solid lines in the main panels
correspond to fits of Eq. (S1) discussed in Supplemental Material \cite{SM}.}
\end{center}
\end{figure*}

When applying a finite $V_\text{th}$, $V(t)$ will also have excursions 
below $V_\text{th}$ (see again Fig. \ref{fig:1}), with  corresponding
time intervals $T_\text{W}$ referred to as the waiting times. In the
simplest possible scaling picture, the excursions above and below
$V_\text{th}$ would have the same statistical properties up to a cutoff
scale. Such a symmetry applies 
in the scaling regime of memory-less Markovian processes, such as simple random 
walks \cite{LAU-09}, but the same may also be true for critical avalanches due to
their self-affine properties. 
The visually asymmetric appearance of the 
$V(t)$ signals with respect to $V_\text{th}$ (Fig. \ref{fig:1}) can be
understood
by noticing that the cutoff mechanisms acting on excursions 
above and below $V_\text{th}$ are different: 
the stiffness parameter $K$ 
(or, e.g., the demagnetizing factor in the case of bursty dynamics of domain 
walls) results in a ``soft'' cutoff mechanism that
limits the growth of $V(t)$ above $V_\text{th}$, giving rise to a cutoff
avalanche duration $T_0 \propto K^{-1/\sigma_K}$. The constraint $V(t) \geq 0$ 
acts as a ``hard'' cutoff for excursions of $V(t)$ below $V_\text{th}$, leading to a 
cutoff waiting time $T_{\text{W},0}$. In the scaling regime, i.e., for $T \ll T_0$ and 
$T_\text{W} \ll T_{\text{W},0}$, we expect  
the statistical properties of the waiting times $T_\text{W}$ to be similar 
to avalanche durations $T$, that is, $P(T_\text{W})$ should be a power 
law with a cutoff
\begin{equation}
\label{eq:PTW}
P(T_\text{W}) =
T_\text{W}^{-\tau_{T_\text{W}}} g'\left(\frac{T_\text{W}}{T_{\text{W},0}}\right),
\end{equation}
with $g'(x)$ another scaling function. Due to the conjectured symmetry
between the excursions of $V(t)$ above and below $V_\text{th}$,
$\tau_{T_\text{W}} = \tau_{T}$.
The boundary condition at $V(t)=0$, together with the symmetry of the excursions
above and below $V_\text{th}$, 
leads to a cutoff waiting time 
$T_{\text{W},0}$ obeying $V_\text{th} \propto T_{\text{W},0}^{\gamma-1}$. Thus
$T_{\text{W},0} \propto V_\text{th}^{\delta}$,
where $\delta = 1/(\gamma-1)$. Since usually $\gamma>1$, 
$T_{\text{W},0}$ thus increases with rising $V_\text{th}$. 
These predictions 
originate from the hypothesis that empirical observations of power law waiting time 
distributions are due to avalanches partly hiding below the detection threshold.
Next, we proceed to test these predictions for experimental and numerical data on 
bursty crack propagation in disordered solids.

In the experiments, a crack is forced to propagate along a heterogeneous
weak plane of a transparent poly(methyl methacrylate) (PMMA) block with an 
imposed constant velocity $\langle V \rangle$ in quasi-mode I geometry \cite{LAU-13,SCH-97,MAL-06}.
A high-resolution fast camera mounted on a microscope 
directly observes the interfacial crack growth (right inset of Fig. \ref{fig:1}). 
The measured crackling noise (top left inset of Fig. \ref{fig:1}) 
corresponds to the time evolution $V(t)$ of the spatially averaged crack front velocity; 
it has been shown to display intermittent avalanche dynamics 
with complex spatio-temporal inter-events correlations \cite{LAU-13,GRO-09,SCH-97,TAL-16}.
For more details, see Supplemental Material \cite{SM}.

The large scale dynamics of our planar crack experiment  
can be described by a model of a long-range elastic, one dimensional
(1D) string propagating in a 2D random medium \cite{LAU-13,LAU-10,BON-08,TAN-98}. 
Here, we perform an extensive set of simulations of its discretized version,
known to capture the avalanche statistics
of the corresponding continuous model \cite{LAU-13,LAU-10,BON-08}, and
represented by a set of integer heights $h_i(t)$, $i = 1 \dots L$, with $L$ the
system size. 
The lateral coordinates $x_i$ of the interface are given by $x_i = i$. The total 
force acting on the interface element $i$ is
\begin{equation}
F_i = \Gamma_\textrm{0}
\sum_{j \neq i}\frac{h_j-h_i}{|x_j-x_i|^2}
+ \eta(x_i,h_i) + F_{\textrm{ext}},
\label{eq:eom}
\end{equation}
where the first term on the right hand side 
represents the long-range elastic
interactions, $\eta$ is uncorrelated quenched disorder modeling toughness
fluctuations of the disordered weak plane, and
$F_{\textrm{ext}}$ is the external driving force.
In addition to planar crack front propagation 
\cite{BON-08, LAU-10, LAU-13}, the model also describes
contact lines of liquids spreading on solid surfaces \cite{ERT-94,JOA-84}
and low-angle grain boundaries in plastically deforming crystals
\cite{MOR-04}. The crackling noise signal is given by
$V(t) = 1/L\sum_i v_i(t)$, where $v_i = \theta (F_i)$, with $\theta$
the Heaviside step function. The interface is driven with a constant
velocity $\langle V \rangle$, by imposing 
$F_{\textrm{ext}}=K(\langle V \rangle t-\langle h \rangle)$, where $K$ describes the
stiffness of the specimen-machine system and controls the cutoffs
$S_0$ and $T_0$, and $\langle h \rangle$ is the average interface height.
For additional details, see Supplemental Material \cite{SM}.

First, we consider the scaling of the average avalanche size
$\langle S(T) \rangle$ with the avalanche duration $T$ for different threshold
levels $V_\text{th}$.
Fig. \ref{fig:2}a, in which experimental data with $\langle V \rangle$ = 10.2 
$\mu$m/s is considered, shows that the effective $\gamma$-value depends on 
$V_\text{th}$ (top left inset of Fig. \ref{fig:2}a, and Supplemental Material, 
Fig. S1 \cite{SM}); the theoretically expected value, $\gamma \approx 1.8$ \cite{LAU-13}, 
is recovered only in the limit $V_\text{th} \ll \langle V \rangle$, while larger $V_\text{th}$-values
lead to smaller effective values of $\gamma$. In particular, using a $V_\text{th}$
maximizing the number of events (this happens for $V_\text{th} \approx \langle V 
\rangle$, a typical choice in experiments, see the bottom right inset of 
Fig. \ref{fig:2}a) would lead to
a $\gamma$-value different from the one obtained in the low-threshold limit.
Fig. \ref{fig:2}b shows that the threshold dependence of the $\langle s(T)       
\rangle$ scaling observed for the experimental data is captured by the model.

\begin{figure}[t!]
\includegraphics[trim=0.0cm 0.0cm 0cm 0.0cm, clip=true, width=7.5cm]{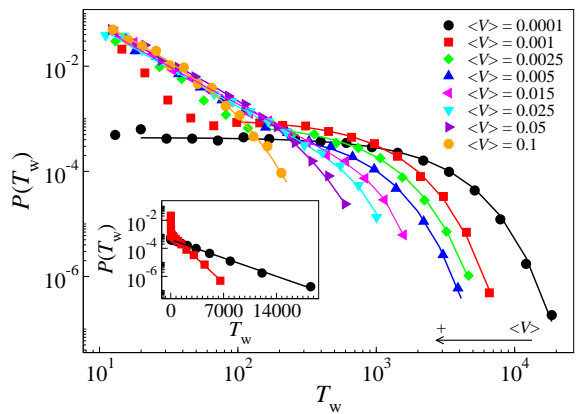}
\caption{The numerically simulated $P(T_\text{W})$'s for a wide
range of 
$\langle V \rangle$, with $V_\text{th}=\langle V \rangle$.
Upon decreasing $\langle V \rangle$ (and thus $V_\text{th}$), $P(T_\text{W})$
evolves from a power law with a cutoff towards a purely exponential distribution
(see the inset for the two distributions with the smallest $\langle V \rangle$
with a semilog axis scale), indicating the absence of correlation in the limit
$\langle V \rangle$, $V_\text{th} \rightarrow 0$. Solid lines in the main panel
correspond to fits of Eq. (S1) discussed in the Supplemental Material \cite{SM}.
\label{fig:4}}
\end{figure}

Next, we present how power-law distributed
waiting times in crackling noise emerge from thresholding. To this end,
Figs. \ref{fig:3}a and b show examples of the experimental $P(T)$ and 
$P(T_\text{W})$ distributions, respectively, for a wide range of threshold 
levels $V_\text{th}$. The (sub)avalanche duration distributions $P(T)$ 
display a power-law terminated at a cutoff $T_0$, with the latter decreasing 
with increasing $V_\text{th}$. Also the $P(T_\text{W})$s display a power law 
with a cutoff, exhibiting the opposite trend to $P(T)$ distributions in that the 
cutoff scale $T_{\text{W},0}$ increases with $V_\text{th}$ as
$T_{\text{W},0} \propto V_\text{th}^{\delta}$, with $\delta = 1.30 \pm 0.10$ (top 
right inset of Fig. \ref{fig:3}b); 
this corresponds to $\gamma \approx 1.77 \pm 0.06$, that is, close to the low-threshold 
result quoted above from the $\langle S(T) \rangle$ scaling. Notably, for 
very small $V_\text{th}$, $P(T_\text{W})$ ceases to have a power-law part 
and is instead close to a pure exponential (bottom left inset of Fig. 
\ref{fig:3}b), indicating that the ``true'' avalanche triggers would be 
well-described by an uncorrelated Poisson process \cite{JAG-14}. Upon increasing 
$V_\text{th}$, avalanches more frequently break into sub-avalanches,
and a power-law part emerges, characterized by an exponent $\tau_{T_\text{W}} 
\approx \tau_{T} = 1.52 \pm 0.05$, 
signaling the onset of apparent correlations due to 
thresholding. 
These results can be reproduced in experiments with 
other $\langle V \rangle$-values (Supplemental Material, Figs. S2 and S3 \cite{SM}).

Figs. \ref{fig:3}c and d show examples of the corresponding numerical 
$P(T)$ and $P(T_\text{W})$ distributions. 
We observe an excellent agreement between simulation and experimental 
results, with $P(T_\text{W})$ evolving from an 
exponential to a power law with increasing $V_\text{th}$. The exponent 
$\tau_{T_\text{W}}$ equals $\tau_T = 1.52 \pm 0.03$ within error bars, and 
$T_{\text{W},0}$ increases with $V_\text{th}$ as $T_{\text{W},0} 
\propto V_\text{th}^{1.3}$ (filled symbols in the top right inset of Fig. 
\ref{fig:3}b). Also the areas $S$ and $S'=\int_0^{T_\text{W}}dt[V_\text{th}-V(t)]$ of 
the excursions of $V(t)$ above and below $V_\text{th}$, respectively, scale with
the same exponent $\tau_S = \tau_{S'}\approx 1.28$ (Supplemental Material, Fig. S4 
\cite{SM}). We also note that a mean field version of Eq. \ref{eq:eom} 
agrees with our scaling picture (Supplemental Material, Figs. S5 and S6 \cite{SM}).

This excellent agreement between experiment and model allows us to apply the latter
to probe the quasistatic limit $\langle V \rangle \rightarrow 0$, not easily
reachable experimentally. Fig. \ref{fig:4} shows the simulated
$P(T_\text{W})$ distributions for a wide range of $\langle V \rangle$-values,
setting $V_\text{th}=\langle V \rangle$. When $V_\text{th} = \langle V \rangle \rightarrow 0$, 
$P(T_\text{W})$ becomes an exponential with a long characteristic
waiting time, and evolves towards a power law with increasing $V_\text{th}=\langle V \rangle$. 
This provides an additional way of looking at how the thresholding process results 
in a power-law $P(T_\text{W})$, even if the underlying ``true'' avalanches are 
triggered by a Poisson process.

Our results show that when bursty events are extracted from a crackling
noise signal by thresholding, they tend to exhibit apparent temporal correlations
visible as power-law distributed waiting times. 
While noise-filtering techniques \cite{PAP-11} may be applied to reduce the
need of thresholding of $V(t)$ signals suffering from experimental noise, finite
sensitivity of any real measurement should lead to a similar outcome. 
This viewpoint agrees with the fact that a large fraction of empirical crackling
noise signals exhibit power law waiting time distributions. Indeed, 
we expect our arguments to be generally applicable 
for any system exhibiting crackling noise, ranging from Barkhausen noise in 
ferromagnets to earthquakes.
The importance of seismic activity below the detection threshold for earthquake
statistics has been discussed by suggesting that small, undetectable shocks may 
trigger detectable events \cite{SOR-05}. Our interpretation would, however, be 
more far-reaching, suggesting that temporally correlated (or clustered) events 
may be parts of the same avalanche. Our work shows how to test this 
{\it a posteriori} by varying the threshold applied to define the crackling noise 
events. When performing such tests, one should bear in mind that strong 
enough additive white noise in the $V(t)$ signal (e.g., due to noisy experimental 
apparatus) is expected to result in scaling properties of the waiting times 
different from the ones reported here \cite{LAU-09}. Our crack propagation
experiments have the advantage of very low levels of experimental noise, and 
thus our experimental results adhere to the noise-free scaling picture of
thresholding-induced waiting times.

Moreover, other processes may be operating in some systems \cite{JAG-14}
in parallel with the thresholding-induced event clustering, and are likely to 
lead to different types of correlations not fully accounted for by our scaling 
description; an interesting possibility would be to modify our experiment
-- by changing the material and/or the experimental conditions -- to add viscoelastic
response to the system. Thus, our work calls for detailed analysis of experimental 
data in diverse crackling noise systems to decipher the origin and nature of 
inter-event correlations or avalanche clustering in each case.

\begin{acknowledgments}
This research has been supported by the Academy of Finland
through an Academy Research Fellowship (L.L., project no. 268302) and
through the Centres of Excellence Program (project no. 251748).
K.J.M. acknowledges the support of the Norwegian Research Council
through the Frinat Grant No. 205486. L.L. wishes to thank S.S. and CNRS
for the hospitality during the invited researcher visit at ENS Lyon.
We acknowledge the computational resources provided by the Aalto University
School of Science ``Science-IT'' project.
\end{acknowledgments}

\end{document}